\documentclass{llncs}

\usepackage[latin1]{inputenc}
\usepackage{graphicx}
\usepackage{amsmath}
\usepackage{mathabx}
\usepackage{url}

\begin{document}

\title{Bounded Model Checking of Temporal Formulas with Alloy\thanks{This work is financed by the ERDF -- European Regional Development Fund
through the COMPETE Programme (operational programme for
competitiveness) and by national funds through the FCT -- Fundação para
a Ciência e a Tecnologia (Portuguese Foundation for Science and
Technology) within project FCOMP-01-0124-FEDER-037281.}}
\author{Alcino Cunha}
\institute{{HASLab} -- High Assurance Software Laboratory\\INESC TEC \& Universidade do Minho, Braga, Portugal\\
\email{alcino@di.uminho.pt}}

\newcommand{\lsem}{\ldbrack}
\newcommand{\rsem}{\rdbrack}

\bibliographystyle{plain}

\maketitle

\begin{abstract}
  Alloy is a formal modeling language based on first-order relational
  logic, with no native support for specifying reactive systems. We
  propose an extension of Alloy to allow the specification of temporal
  formulas using LTL, and show how they can be verified by bounded
  model checking with the Alloy Analyzer.  
\end{abstract}

\section{Introduction}

Alloy is a formal modeling language based on first-order relational logic~\cite{DBLP:books/daglib/0024034}.  Its Analyzer enables model
validation and verification by translation to off-the-shelf SAT
solvers. Alloy's logic is quite generic and does not commit to a
particular specification style. In particular, there is no predefined
way to specify and verify reactive systems, and several idioms and
extensions have been proposed to address this issue. However, it is
rather cumbersome and error-prone to specify and verify temporal
properties with such idioms. In this paper we propose the usage of
standard \emph{Linear Temporal Logic} (LTL) to specify reactive systems in Alloy, and show how
bounded model checking can be performed with its Analyzer,
by resorting to the technique first proposed
by Biere et al~\cite{DBLP:conf/tacas/BiereCCZ99}.

This paper is structured as follows. Section~\ref{sec:model-trac-spec}
shows how reactive systems and temporal properties can be specified
and verified using Alloy and its Analyzer. Section~\ref{sec:ltl}
discusses how such properties can be specified more easily in LTL and
then translated to Alloy, avoiding some of the potential problems
pointed out in the previous section. Finally, we discuss some related
work in Section~\ref{sec:related-work}.

\section{Verifying reactive systems in Alloy}
\label{sec:model-trac-spec}

In Alloy, a {\em signature} represents a set of atoms.  An atom is a
unity with three fundamental properties: it is indivisible, immutable
and uninterpreted. A signature declaration can introduce
\emph{fields}, sets of tuples of atoms capturing \emph{relations}
between the enclosing signature and others. Model constraints are defined by {\em facts}. {\em
  Assertions} express properties that are expected to hold as
consequence of the stated facts.  {\em Commands} are instructions to
perform particular analysis.  Alloy provides two commands: \texttt{run}, that instructs the Analyzer to search for an
instance satisfying a given formula, and \texttt{check}, that attempts
to contradict a formula by searching for a counterexample.

Since fields are immutable, to capture the dynamics of a state transition system a special signature
whose atoms denote the possible states must be
declared. Without loss of generality, we will denote this
signature as \texttt{State}. Every field specifying a
mutable relation must then include \texttt{State} as one of the
signatures it relates. There are two typical Alloy \emph{idioms} to
declare such mutable fields: declare them all inside \texttt{State},
or add \texttt{State} as the last column in every mutable field
declaration. The former idiom is known as \emph{global state}, since
all mutable fields are grouped together, while the latter is known as
\emph{local state}, since mutable fields are declared in the same
signature as non-mutable ones of similar type.


\begin{figure}[t]
{\small
\begin{verbatim}
open util/ordering[State]

sig State     {}
sig Message   { to, from : one Partition }
sig Channel   { messages : Message set -> State }
sig Partition { port : one Channel, pifp : set Partition }

fact NoSharedChannels { all c : Channel | lone port.c }

pred send     [m : Message, s,s' : State] { ... }
pred receive  [m : Message, s,s' : State] { ... }
pred transfer [m : Message, s,s' : State] { 
  m in m.from.port.messages.s and m.to in (m.from).pifp
  messages.s' = messages.s - m.from.port->m + m.to.port->m
}

fact Init  { no Channel.messages.first }
fact Trans { all s : State, s' : s.next | 
  some m : Message | send[m,s,s'] or receive[m,s,s'] or transfer[m,s,s']
}
\end{verbatim}
}
\caption{A PIFP specification in Alloy.}
\label{fig:pifp}       
\end{figure} 

Figure~\ref{fig:pifp} presents an example of an Alloy model conforming
to the local state idiom. It is a simplified model of the
\emph{Partition Information Flow Policy} (PIFP) of a \emph{Secure
  Partitioning Kernel}.  Essentially, the PIFP
statically defines which information flows (triggered by message
passing) are authorized between partitions.  Signature
\texttt{Message} declares the fields \texttt{to} and \texttt{fom},
capturing its destination and source \texttt{Partition},
respectively. Communication is done via channels, here simplified to
contain sets of messages. Obviously, the messages contained in a channel vary
over time. As such, signature \texttt{Channel} declares a mutable
relation \texttt{messages}, that associates each channel with
the set of its messages in each state. Signature
\texttt{Partition} declares two binary relations: \texttt{port},
denoting the channel used by the partition to communicate, and
\texttt{pifp}, that captures to which partitions it is authorized to send
messages.

In Alloy \emph{everything is a relation}. For example, sets are unary relations and variables are just unary singleton
relations. As such, relational operators (in particular the \emph{dot join} composition) can be used
for various purposes. For example, in the fact
\texttt{NoSharedChannels} the relational expression \texttt{port.c}
denotes the set of partitions connected to port \texttt{c}. Multiplicities are
also used in several contexts to constrain or check the cardinality of
a relation. For example, in the same fact,
multiplicity \texttt{lone} ensures that each channel is the port of at
most one partition. 


An operation can be specified using a predicate \texttt{pred
  op[...,s,s':State]} specifying when does \texttt{op} hold between the
given pre- and post-states: to access the value of a mutable field in
the pre-state (or post-state) it suffices to compose it with
\texttt{s} (respectively \texttt{s'}). Our model declares three
operations: \texttt{send} and \texttt{receive} message (whose specifications are omitted due to space limitations), and
\texttt{transfer} message between partitions. The \texttt{send}
operation just deposits the message in the sending partition port -- the
\texttt{transfer} operation is executed by the kernel and is the one
responsible to enforce the PIFP. Inside an operation, a formula that
does not refer \texttt{s'} can be seen as a pre-condition. Otherwise
it is a post-condition. For example, \texttt{m in
  m.from.port.messages.s} is a pre-condition to \texttt{transfer},
requiring \texttt{m} to be in the port of the source
partition prior to its execution.  Notice that frame conditions, specifying which mutable
relations remain unchanged, should be stated explicitly in each
operation.

To specify temporal properties we need to model execution traces. A
typical Alloy idiom for representing finite prefixes of traces is to impose a total ordering on signature \texttt{State} (by
including the parameterized module \texttt{util/ordering}) and force every pair of
consecutive states to be related by one of the operations (see fact
\texttt{Trans}). Inside module
\texttt{util/ordering}, the total order is defined by the binary relation
\texttt{next}, together with its
\texttt{first} and \texttt{last} states. The initial state of our example, where all channels are empty, is constrained by fact
\texttt{Init}. A desirable safety property states that the port of
every partition only contains messages sent from authorized partitions:

{\small
\begin{verbatim}
assert Safety { all p : Partition, m : Message | all s : State |
  m.to = p and m in p.port.messages.s implies p in m.from.pifp
}
\end{verbatim}
}

Model checking this assertion with the Alloy Analyzer
yields a counter-example: every partition can send a message to itself,
even if not allowed by the PIFP. To correct this problem we can, for
example, add to our model the fact  \texttt{all p:Partition | p in p.pifp}, stating that all partitions should be allowed to send messages to themselves.
Non-invariant temporal assertions can also be expressed with this
idiom, but the complexity of the formulas and expertise required by
the modeler increases substantially. Consider, for example the
liveness property stating that all authorized messages are eventually
transferred to the destination. At first glance, it could
be specified as follows (using transitive closure to access the successors of a given state):

{\small
\begin{verbatim}
assert Liveness { all p : Partition, m : Message | 
  all s : State | m in p.port.messages.s and m.to in p.pifp implies 
    some s' : s.*next | m in m.to.port.messages.s'
}
\end{verbatim}}

A simple (but artificial) way to ensure \texttt{Liveness} in this
model is to disallow message sending while there are still pending
messages to transfer. This could be done by adding the
pre-condition \texttt{no Channel.messages.s} to operation
\texttt{send}. However, even with such pre-condition, model-checking assertion \texttt{Liveness}
with the Alloy Analyzer yields a false counter-example, where a
message is sent in the last state of the trace prefix. In fact, if we consider
only finite prefixes of execution traces, it is almost always possible
to produce a false counter-example to a liveness property. This
problem is well-known in the bounded model-checking community, and the
solution, first proposed in~\cite{DBLP:conf/tacas/BiereCCZ99}, is to
only consider as (true) counter-examples to such properties prefixes of
traces containing a \emph{back loop} in the last state, i.e., those that
actually model infinite execution traces.  It is easy to define a
parameterized \texttt{trace} module\footnote{Available at \url{http://www.di.uminho.pt/~mac/Publications/trace.als}.}, that adapts
\texttt{util/ordering} to specify potential infinite traces instead of total
orders, by allowing such back loop in the last state. Module \texttt{trace} also defines a predicate
\texttt{infinite} that checks if the loop is
present in a trace instance: if so, \texttt{next} always assigns a successor to
every state, thus modeling an infinite trace. For
convenience a dual predicate \texttt{finite} is also defined.
By replacing \texttt{open util/ordering[State]} with \texttt{open
  trace[State]}, the above liveness property can now be correctly specified
(and verified) as follows:

{\small
\begin{verbatim}
assert Liveness { all p : Partition, m : Message |
  all s : State | m in p.port.messages.s and m.to in p.pifp implies 
    finite or some s' : s.*next | m in m.to.port.messages.s'
}
\end{verbatim}
}

\section{Embedding LTL formulas in Alloy}
\label{sec:ltl}

As seen in the previous section, although we can specify and
verify (by bounded model checking) temporal properties in standard Alloy, it is a rather
tricky and error-prone task, in particular since the user must be careful about
where to check for finitude of trace prefixes. As such, we propose that, instead of using explicit quantifiers over the states in a trace, such
properties be expressed using the standard LTL operators: \texttt{X} for next, \texttt{G} for
always, \texttt{F} for eventually, \texttt{U} for until, and
\texttt{R} for release.
For example, the above temporal properties could be specified as follows:

{\small
\begin{verbatim}
assert Safety { all p : Partition, m : Message |
  G (m.to = p and m in p.port.messages implies p in m.from.pifp)
}
\end{verbatim}
}
\newpage
{\small
\begin{verbatim}
assert Liveness { all p : Partition, m : Message |
  G (m in p.port.messages and m.to in p.pifp implies 
    F (m in m.to.port.messages))
}
\end{verbatim}
}


Assuming traces are specified with module \texttt{trace},
the embedding of LTL into Alloy can be done via an (almost)
direct encoding of the translation proposed
by Biere et al.~\cite{DBLP:conf/tacas/BiereCCZ99} (for bounded model checking of LTL
with a SAT solver). Formally, a formula $\phi$ occurring in a
fact or \texttt{run} command should be replaced by $\lsem
\mathit{NNF}(\phi) \rsem_{\mathtt{first}}$, where $\lsem \phi \rsem_s$
is the embedding function defined in Figure~\ref{fig:ltl2alloy}, and
$\mathit{NNF}(\phi)$ is the well-known transformation that converts
formula $\phi$ to \emph{Negation Normal Form} (where all negations
appear only in front of atomic formulas). When finding a model for
$\mathtt{G}\ \phi$, only prefixes capturing infinite traces should be
considered, thus assuring that $\phi$ is not violated
further down the trace. As clarified in~\cite{DBLP:conf/tacas/BiereCCZ99}, conversion to NNF is necessary since in the bounded semantics of LTL the duality of \texttt{G} and \texttt{F} no longer hold. In the encoding of \texttt{U} and \texttt{R} we use the function \texttt{upto}, defined in module \texttt{trace}, that, given $s$ and $s'$ computes all states from $s$ up to $s'$ (not including the latter).
The embedding of logic and relational
operators is trivial, and thus only a representative subset of Alloy's logic is presented.  A formula $\phi$ occurring in an assertion or
\texttt{check} command should be replaced by $\mathtt{not}\ \lsem
\mathit{NNF}(\mathtt{not}\ \phi) \rsem_{\mathtt{first}}$. Since
assertions in check commands are negated in order to find counter-examples,
the outermost negation ensures they still remain in NNF.

\begin{figure}[t]
  \centering
  \begin{displaymath}
    \begin{array}{c}
  \begin{array}{rl}
    \lsem \mathtt{X}\ \phi \rsem_s \equiv & \mathtt{some}\ s\mathtt{.next}\ \mathtt{and}\ \lsem \phi \rsem_{s\mathtt{.next}}\\
    \lsem \mathtt{G}\ \phi \rsem_s \equiv & \mathtt{infinite\ and\ all}\ s'\, \mathtt{:}\, s\texttt{.*next}\ \mathtt{|}\ \lsem \phi \rsem_{s'}\\
    \lsem \mathtt{F}\ \phi \rsem_s \equiv & \mathtt{some}\ s'\, \mathtt{:}\, s\texttt{.*next}\ \mathtt{|}\ \lsem \phi \rsem_{s'}\\
    \lsem \phi\ \mathtt{U}\ \psi \rsem_s \equiv & \mathtt{some}\ s'\, \mathtt{:}\, s\texttt{.*next}\ \mathtt{|}\ \lsem \psi \rsem_{s'}\, \mathtt{and}\ \mathtt{all}\ s''\, \mathtt{:}\, \texttt{upto[}s\texttt{,}s'\texttt{]}\ \mathtt{|}\ \lsem \phi \rsem_{s''}\\
    \lsem \phi\ \mathtt{R}\ \psi \rsem_s \equiv & \lsem \mathtt{G}\ \psi \rsem_s\ \mathtt{or}\ \mathtt{some}\ s'\, \mathtt{:}\, s\texttt{.*next}\ \mathtt{|}\ \lsem \phi \rsem_{s'}\, \mathtt{and}\ \mathtt{all}\ s''\, \mathtt{:}\, \texttt{upto[}s\texttt{,}s'\texttt{]}\texttt{+}s'\ \mathtt{|}\ \lsem \psi \rsem_{s''}
   \end{array}\\\\
   \begin{array}{rl@{\qquad}rl}
     \lsem \mathtt{not}\ \phi \rsem_s \equiv & \mathtt{not}\ \lsem \phi \rsem_s & \lsem \Phi\ \texttt{.}\ \Psi \rsem_s \equiv & \lsem \Phi \rsem_s\ \texttt{.}\  \lsem \Psi \rsem_s\\
     \lsem \phi\ \mathtt{and}\ \psi \rsem_s \equiv & \lsem \phi \rsem_s\ \mathtt{and}\  \lsem \psi \rsem_s & \lsem \Phi\ \texttt{\&}\ \Psi \rsem_s \equiv & \lsem \Phi \rsem_s\ \texttt{\&}\  \lsem \Psi \rsem_s\\
     \lsem \phi\ \mathtt{or}\ \psi \rsem_s \equiv & \lsem \phi \rsem_s\ \mathtt{or}\  \lsem \psi \rsem_s & \lsem \Phi\ \texttt{+}\ \Psi \rsem_s \equiv & \lsem \Phi \rsem_s\ \texttt{+}\  \lsem \Psi \rsem_s\\
     \lsem \mathtt{all}\ x\ \mathtt{:}\ \Phi\ \mathtt{|}\ \phi \rsem_s \equiv & \mathtt{all}\ x\ \mathtt{:}\ \lsem \Phi \rsem_s\ \mathtt{|}\ \lsem \phi \rsem_s & \lsem \Phi\ \texttt{->}\ \Psi \rsem_s \equiv & \lsem \Phi \rsem_s\ \texttt{->}\  \lsem \Psi \rsem_s\\
     \lsem \mathtt{some}\ x\ \mathtt{:}\ \Phi\ \mathtt{|}\ \phi \rsem_s \equiv & \mathtt{some}\ x\ \mathtt{:}\ \lsem \Phi \rsem_s\ \mathtt{|}\ \lsem \phi \rsem_s & \lsem \texttt{*} \Phi \rsem_s \equiv & \texttt{*}  \lsem \Phi \rsem_s \\
    \lsem \Phi\ \texttt{in}\ \Psi \rsem_s \equiv & \lsem \Phi \rsem_s\ \texttt{in}\  \lsem \Psi \rsem_s & \lsem \mathtt{none} \rsem_s \equiv & \mathtt{none}
     \end{array}\\\\
   \lsem x \rsem_s \equiv \left\{
     \begin{array}{l@{\quad}l}
       x\mathtt{.}s & \text{if $x$ is the id of a mutable field declared with the local state idiom}\\
       s\mathtt{.}x & \text{if $x$ is the id of a mutable field declared with the global state idiom}\\
       x & \text{otherwise (i.e., a variable or the id of an immutable field)}
     \end{array}
     \right.
   \end{array}  
\end{displaymath}
   \caption{Embedding of temporal formulas.}
   \label{fig:ltl2alloy}
\end{figure}

Note that, to improve efficiency (and likewise to \texttt{util/ordering}), when the \texttt{trace} module is
imported the scope of the parameter signature is interpreted as an
exact scope. This means that trace prefixes are bounded to be of size equal to the
scope of the \texttt{State} signature. Thus, to perform bounded model
checking of an assertion, the user should manually
increase the scope of \texttt{State} one unit at a time up to the desired
bound.

\section{Related work}
\label{sec:related-work}

Several extensions of Alloy to deal with dynamic behavior have been
proposed. DynAlloy~\cite{DBLP:conf/icse/FriasGPA05} proposes an Alloy
variant that allows the specification of properties over execution
traces using a formalism inspired by dynamic logic.  Imperative
Alloy~\cite{DBLP:conf/asm/NearJ10} proposes a more minimal extension
to the language, with a simple semantics by means of an
embedding to standard Alloy. Unfortunately, in both these works the
verification of liveness properties may yield spurious
counter-examples, similar to the one presented in
Section~\ref{sec:model-trac-spec}.

One of the advantages of our approach is that reactive systems can be
specified declaratively using Alloy's relational logic, as opposed to
traditional model checkers where transitions must be specified
imperatively. Chang and Jackson~\cite{DBLP:conf/icse/ChangJ06}
proposed a BDD-based model checker for declarative models specified with relational logic enhanced with CTL temporal formulas. The current proposal shows how the Alloy Analyzer
can directly be used to perform bounded model checking of temporal
formulas without the need for a new tool.

Recently, Vakili and Day~\cite{DBLP:conf/asm/VakiliD12} showed how CTL
formulas with fairness constraints can be model checked in Alloy, by
using the encoding to first order logic with transitive closure first
proposed by Immerman and Vardi~\cite{DBLP:conf/cav/ImmermanV97}. Their technique performs
full model checking on state transition systems specified
declaratively, but bounded to have at most the number of states
specified in the scope. This non-standard form of bounded model
checking can yield non-intuitive results in many application
scenarios, or even prevent verification at all if the the
specification cannot be satisfied by a transition system that fits in the (necessarily small) scope of \texttt{State}. Moreover, instead of proposing an Alloy extension,
CTL formulas are expressed using library functions that compute the
set of states where the formula holds. This leads to unintuitive
specifications, since the user is then forced to use relational operators
to combine formulas instead of the standard logical connectives.



\bibliography{refs}

\begin{thebibliography}{1}

\bibitem{DBLP:conf/tacas/BiereCCZ99}
Armin Biere, Alessandro Cimatti, Edmund Clarke, and Yunshan Zhu.
\newblock Symbolic model checking without {BDD}s.
\newblock In {\em TACAS}, volume 1579 of {\em LNCS}, pages 193--207. Springer,
  1999.

\bibitem{DBLP:conf/icse/ChangJ06}
Felix Chang and Daniel Jackson.
\newblock Symbolic model checking of declarative relational models.
\newblock In {\em ICSE}, pages 312--320. ACM, 2006.

\bibitem{DBLP:conf/icse/FriasGPA05}
Marcelo Frias, Juan Galeotti, Carlos Pombo, and Nazareno Aguirre.
\newblock {DynAlloy}: upgrading {A}lloy with actions.
\newblock In {\em ICSE}, pages 442--451. ACM, 2005.

\bibitem{DBLP:conf/cav/ImmermanV97}
Neil Immerman and Moshe Vardi.
\newblock Model checking and transitive-closure logic.
\newblock In {\em CAV}, volume 1254 of {\em LNCS}, pages 291--302. Springer,
  1997.

\bibitem{DBLP:books/daglib/0024034}
Daniel Jackson.
\newblock {\em Software Abstractions - Logic, Language, and Analysis}.
\newblock MIT Press, revised edition, 2012.

\bibitem{DBLP:conf/asm/NearJ10}
Joseph Near and Daniel Jackson.
\newblock An imperative extension to {A}lloy.
\newblock In {\em ABZ}, volume 5977 of {\em LNCS}, pages 118--131. Springer,
  2010.

\bibitem{DBLP:conf/asm/VakiliD12}
Amirhossein Vakili and Nancy Day.
\newblock Temporal logic model checking in {A}lloy.
\newblock In {\em ABZ}, volume 7316 of {\em LNCS}, pages 150--163. Springer,
  2012.

\end{thebibliography}

\end{document}